\documentclass[showpacs,floatfix,amssymb,amsmath,aps,prb,twocolumn,eqsecnum]{revtex4}
\usepackage[dvips]{graphicx}
\usepackage[ps2pdf,colorlinks=false,pdfpagemode=UseNone]{hyperref}
\usepackage{latexsym,epsfig,bm,times,psfrag,subfigure}
\usepackage{dcolumn}
\begin{document}

\newcommand{\dos}{D}
\newcommand{\curO}{{\cal O}}
\renewcommand{\r}{{\bf{r}}}
\renewcommand{\S}{{\mathcal{S}}}
\renewcommand{\k}{{\bf{k}}}
\renewcommand{\arcsin}{\operatorname{arc\,sin}}
\renewcommand{\arccos}{\operatorname{arc\,cos}}
\newcommand{\arctanh}{\operatorname{arc\,tanh}}
\newcommand{\as}{a_s}
\newcommand{\g}{g}
\newcommand{\phdagg}{^{\phantom{\dagger}}}
\newcommand{\phid}{\phi^{\dagger}}
\newcommand{\phipd}{\phi}
\newcommand{\psid}{\psi^{\dagger}}
\newcommand{\psipd}{\psi^{\phantom{\dagger}}}
\newcommand{\phis}{\phi^{\ast}}
\newcommand{\phips}{\phi^{\phantom{\ast}}}
\newcommand{\psis}{{\psi^{\ast}}}
\newcommand{\psips}{\psi^{\phantom{\ast}}}
\renewcommand{\psips}{\psi}
\renewcommand{\psipd}{\psi}
\renewcommand{\phips}{\phi}
\renewcommand{\phipd}{\phi}
\newcommand{\phio}{\overline{\phi}}
\newcommand{\phipo}{{\phi}}
\newcommand{\psio}{\overline{\psi}}
\renewcommand{\phipo}{{\phi}}
\renewcommand{\phio}{\phi^\ast}
\newcommand{\psipo}{{\psi}}
\newcommand{\etao}{\overline{\eta}}
\newcommand{\etapo}{{\eta}}
\newcommand{\Psio}{\overline{\Psi}}
\newcommand{\Psipo}{{\Psi}}
\newcommand{\Etao}{\overline{\Phi}}
\newcommand{\Etapo}{{\Phi}}
\newcommand{\order}[1]{\mathcal{O}(#1)}
\newcommand{\mb}[1]{\mathbf{#1}}
\newcommand{\bo }{\Delta}
\renewcommand{\b}{\hat{b}}
\newcommand{\be}{\begin{equation}}
\newcommand{\ee}{\end{equation}}
\newcommand{\bea}{\begin{eqnarray}}
\newcommand{\eea}{\end{eqnarray}}
\newcommand{\ba}{\begin{align}}
\newcommand{\ea}{\end{align}}
\newcommand{\s}{{\sigma}}
\newcommand{\si}{{\sigma^\prime}}
\newcommand{\lamda}{\lambda}
\newcommand{\eps}{\epsilon}
\newcommand{\rp}{{\bf{r}^{\prime}}}
\newcommand{\rf}{{\bf{r}}_f}
\newcommand{\ri}{{\bf{r}}_i}
\newcommand{\q}{{\bf{q}}}
\newcommand{\qp}{{\bf{q^{\prime}}}}
\newcommand{\kp}{{\bf{k^{\prime}}}}
\newcommand{\p}{{\bf{p}}}
\newcommand{\pp}{{\bf{p^{\prime}}}}
\newcommand{\ham}{\mathcal{H}}
\newcommand{\G}{{\mathcal{G}}}
\newcommand{\GF}{{\mathcal{\tilde{G}}}}
\newcommand{\epsl}{\epsilon}
\newcommand{\uu}{\uparrow}
\newcommand{\dd}{\downarrow}
\newcommand{\Tr}{{\rm{Tr}}}
\newcommand{\orot}{\Omega_{r}}
\renewcommand{\orot}{\omega}
\newcommand{\op}{\Omega_{\perp}}
\newcommand{\oz}{\Omega_{z}}
\newcommand{\barg}{\overline{g}}
\newcommand{\sech}{\operatorname{sech}}
\newcommand{\espilon}{\epsilon}
\newcommand{\OT}{\overline{\Omega}}
\newcommand{\OTO}{\tilde{\Omega}_0}
\newcommand{\az}{{a^{z}}}
\newcommand{\xb}{{\overline{x}}}
\newcommand{\yb}{{\overline{y}}}
\newcommand{\zb}{{\overline{z}}}
\newcommand{\zo}{{z_0}}
\newcommand{\zp}{{z_+}}
\newcommand{\zm}{{z_-}}
\newcommand{\Qo}{{Q^{\dagger}_0}}
\newcommand{\Qp}{{Q^{\dagger}_+}}
\newcommand{\Qm}{{Q^{\dagger}_-}}
\newcommand{\Qpo}{{Q^{\phantom{\dagger}}_{0}}}
\newcommand{\Qpp}{{Q^{\phantom{\dagger}}_{+}}}
\newcommand{\Qpm}{{Q^{\phantom{\dagger}}_{-}}}
\newcommand{\epp}{\overline{\epsilon_0}}
\newcommand{\eF}{{\epsilon}_{F}}
\newcommand{\taup}{\tau^{\prime}}
\newcommand{\numba}{-0.163} 
\newcommand{\numbb}{-0.312} 
\newcommand{\numbc}{-1.31} 
\newcommand{\numbd}{-0.58} 
\newcommand{\numbe}{+0.087} 
\newcommand{\numbf}{-0.631}  
\newcommand{\numbg}{-0.458} 
\newcommand{\numbh}{-0.242}   
\newcommand{\numbi}{*19}
\newcommand{\numbj}{*10}
\newcommand{\NS}{N}
\newcommand{\NNN}{n}
\newcommand{\SSS}{s}
\newcommand{\SNS}{S}
\newcommand{\SOO}{S^{(0)}}
\newcommand{\STT}{S^{(1/\NS)}}
\newcommand{\EEE}{\varepsilon}
\newcommand{\MEANE}{\epsilon}
\newcommand{\FFF}{f}
\newcommand{\FFS}{\FFF_\SSS}
\newcommand{\FFN}{\FFF_\NNN}
\newcommand{\EES}{\EEE_\SSS}
\newcommand{\EEN}{\EEE_\NNN}
\newcommand{\EET}{\EEE^{(1/\NS)}}
\newcommand{\FFT}{\FFF^{(1/\NS)}}
\newcommand{\EEO}{\EEE^{(0)}}
\newcommand{\FFO}{\FFF^{(0)}}
\newcommand{\EEST}{\EEE^{(1/\NS)}_\SSS}
\newcommand{\FFST}{\FFF^{(1/\NS)}_\SSS}
\newcommand{\EESO}{\EEE^{(0)}_\SSS}
\newcommand{\FFSO}{\FFF^{(0)}_\SSS}
\newcommand{\EENT}{\EEE^{(1/\NS)}_\NNN}
\newcommand{\FFNT}{\FFF^{(1/\NS)}_\NNN}
\newcommand{\EENO}{\EEE^{(0)}_\NNN}
\newcommand{\FFNO}{\FFF^{(0)}_\NNN}
\newcommand{\qat}{^{(0)}_c}
\newcommand{\qao}{^{(0)}_o}
\newcommand{\qah}{^{(0)}_{c2}}
\renewcommand{\log}{\ln}
\title{Large-$N$ expansion for unitary superfluid Fermi gases}
\author{ Martin Y.\ Veillette}
\author{Daniel E.\ Sheehy}
\altaffiliation[Present address: ]{Ames Laboratory and Department of Physics, Iowa State University, Ames, IA 50011}
\author{Leo Radzihovsky}
\affiliation{Department of Physics, University of Colorado, Boulder,
 Colorado 80309}
\date{\today}

\pacs{03.75.Fi, 03.75.Lm, 05.30.Jp}
\begin{abstract}

  We analyze strongly interacting Fermi gases in the unitary regime by
  considering the generalization to an arbitrary number $\NS$ of
  spin-$1/2$ fermion flavors with $Sp(2\NS)$ symmetry. For $\NS\to
  \infty$ this problem is exactly solved by the BCS-BEC
  mean-field theory, with corrections small in the parameter
  $1/\NS$.  The large-$\NS$ expansion provides a systematic way to
  determine corrections to mean-field predictions, allowing the
  calculation of a variety of thermodynamic quantities at (and in
  proximity to) unitarity, including the energy, the pairing gap, and
  the upper-critical polarization (in the case of a polarized gas) for the
  normal to superfluid instability.  For the physical case of $\NS=1$,
  among other quantities, we predict in the unitarity regime, the
  energy of the gas to be $\xi=0.28$ times that for the
  non-interacting gas and the pairing gap to be $0.52$ times the Fermi
  energy.

\end{abstract}
\maketitle

\section{Introduction}
\label{Introduction}

\subsection{Motivation and Background}
\label{Motivation}

Recent advances in atomic
gases near a Feshbach resonance (FR) have led to the experimental
realization of resonantly-paired superfluidity of fermionic atomic
gases~\cite{Regal2004,Zwierlein2004,Kinast2004,Bartenstein2004,Bourdel2004,Partridge2005}.  Although
atomic interactions are generally weak, such superfluidity has been
achieved using the strong attractive interactions provided by
proximity to a magnetic-field tuned Feshbach resonance between the two
hyperfine levels (isomorphic to spin-1/2 states) undergoing pairing.
Indeed, the tunability of such Feshbach resonances, experimentally
controllable via an external magnetic field, allows unprecedented
access to a wide range of fermionic interaction strengths
characterized by the vacuum s-wave scattering length $\as$ that
diverges for a resonance tuned to zero energy.  As the FR detuning
$\delta \propto -1/\as$ is varied, the character of superfluidity
evolves from a weakly-paired Bardeen-Cooper-Schrieffer (BCS) regime at
large positive FR detuning (where $\as<0$) to a strongly-paired
molecular Bose-Einstein Condensate (BEC) regime at large negative
detuning (where $\as>0$).
The experimental signatures of such pairing and superfluidity in
ultracold gases of $^{40}$K and $^6$Li have included direct
measurements of the condensate density~\cite{Regal2004,Zwierlein2004}
and the pairing gap~\cite{Chin2004,Partridge2005} and the observation
of vortex lattices in rotating clouds\cite{Zwierlein2005}.

From a theoretical perspective, a quantitative description of such
resonantly-interacting Fermi superfluids is well-developed away from
the resonance, where $\as$ is short compared to atom spacing. Then,
a controlled perturbative expansion in a natural small parameter
$|\as|n^{1/3}\ll 1$ (with $n$ the fermion density) allows a
quantitative theoretical analysis of both the deep
BCS~\cite{Eagles1969,Leggett1980} and deep
BEC~\cite{Petrov2004,Brodsky2005,Levinsen2005} regimes.
However, the aforementioned present-day experiments are typically in
the crossover between the BCS and BEC regimes, where $|\as|n^{1/3}\gg
1$. Thus, the absence of a small parameter~\cite{narrownote,ONE,GRreview}
near the resonance precludes a systematic perturbative expansion for a
quantitative description of this theoretically challenging regime.

A particular point of interest is the so-called unitarity point,
precisely at zero FR detuning, at which $\as^{-1}=0$ and fermion
scattering is characterized by a maximum scattering phase shift of
$\pi/2$.  At this point, the system does not contain any scale
besides the Fermi wavelength $k_F^{-1}$ set by the atom density
$n=k_F^3/ (3 \pi^2)$, and the only energy scale is the Fermi energy
$\eF=\hbar^2 k_F^2/2m$ and therefore the free energy, and any
quantities related to it, is given by $\eF$ multiplied by a {\it
universal\/} dimensionless function of $k_B T/\eF$ \cite{Ho2004}. In
particular, this implies that, at zero temperature, the internal
energy per particle $\MEANE$ is simply proportional to that of a
non-interacting Fermi gas
\be \MEANE= \xi\ \frac{3}{5} \eF
\label{1.1}. \ee %
Recently, there has been much theoretical interest in computing $\xi$
and other universal parameters, motivated by the possibility of
attaining a quantitative understanding of unitary quantum gases as
realized in cold-atom experiments but also having applications to
nuclear physics and astrophysical systems such as neutron stars
\cite{Baker1999,Heiselberg2001,Perali2004,Astrakharchik2004,Carlson2003,Hu2006,Haussmann2006,Burovski2006,Bulgac2006,Nishida2006a,Nishida2006b,Arnold2006}.
The purpose of this paper is to provide a theoretical framework for a
{\em systematic} determination of the value of such universal
parameters at unitarity.  Our method is based on the introduction of
an artificial small parameter $1/\NS$, with $\NS$ the number of
distinct ``spin''-1/2 fermion flavors with $Sp(2\NS)$ symmetry. We
then use it to extrapolate to the experimentally-relevant case of a
single flavor ($\NS=1$) of two opposite-\lq\lq spin\rq\rq-1/2
(hyperfine levels) fermionic atoms.

The motivation for such a generalization is that, for $\NS \to
\infty$, the problem may be solved exactly~\cite{SachdevWang}, with
the solution taking the form of the standard BCS mean-field theory. We
can then compute corrections in the small parameter $1/\NS$, obtaining
a systematic expansion (in principle to arbitrary order) about this
solvable limit.  Since it is believed that there are no phase
transitions with decreasing $\NS$ (the large-$\NS$ solution having the
same broken symmetry as the exact ground-state at $\NS = 1$), we
expect the large-$\NS$ results to be smoothly connected to the
physical case of $\NS=1$.  The large-$\NS$ expansion may be thought of
as a way to systematically organize corrections around the well-known
BCS mean-field solution. Since this $\NS\to\infty$ solution is known
to give a relatively good estimate, we expect the $1/\NS$-expansion to
converge rapidly.

To further motivate our study, it is instructive to briefly review
other theoretical approaches.  As we have noted, near the unitarity
point a {\em quantitative} theoretical understanding of this strongly
correlated problem is limited by the absence of any physical small
parameter\cite{narrownote,GRreview}.  However, since, the
thermodynamic quantities are expected to evolve smoothly from the BCS
to the BEC limits, a {\em qualitative} description has come from
various uncontrolled (by any small parameter) schemes, that
interpolate between these two regimes.  Starting from a variational
wavefunction (equivalent to mean-field theory), Eagles
\cite{Eagles1969} and Leggett \cite{Leggett1980} studied the crossover
from the BCS to the BEC limit at zero temperature.  Nozieres and
Schmitt-Rink studied the same problem at finite temperature, taking
into account the attraction of fermions beyond mean-field theory
\cite{Nozieres1985}. The repulsion between bound pairs was first
worked out using a functional integral formalism \cite{Sademelo1993,
  Drechsler1992} and a self-consistent theory \cite{Haussmann1993}.
However, except for numerical Monte Carlo calculations
\cite{Carlson2003,Astrakharchik2004,Burovski2006,Bulgac2006}, a full
description of the system around unitarity in a {\em quantitative}
fashion has remained elusive due to the absence of a small parameter.

More recently, stimulated by progress made in the theory of critical
phenomena (where, similarly, no physical small parameter exists near a
critical point)\cite{CriticalPhenomena}, progress has been made by
studying the system in $d$ dimensions, treating the unitary regime in
a systematic expansion in a small parameter associated with the
dimensionality of space\cite{Nussinov2004,Nishida2006a,Nishida2006b,Arnold2006}.

The large-$\NS$ approach used here is close in spirit, but is
complementary to such an expansion-in-dimension study, in that we also
introduce an artificial, but distinct small parameter and perform a
systematic perturbative expansion in it. Such an approach has also
been extremely successfully in a variety of field theory and
statistical physics contexts \cite{Brezin1973,Ma1974,Moshe2003},
applied to a description of continuous phase transitions close to a
critical point.

\subsection{Summary of results}
\label{results}

\begin{figure}[hb]
\begin{center}
\begin{picture}(8.6,10.5)(0,0)
\put(-0.0575,5.2){\includegraphics[width=8.6358cm]{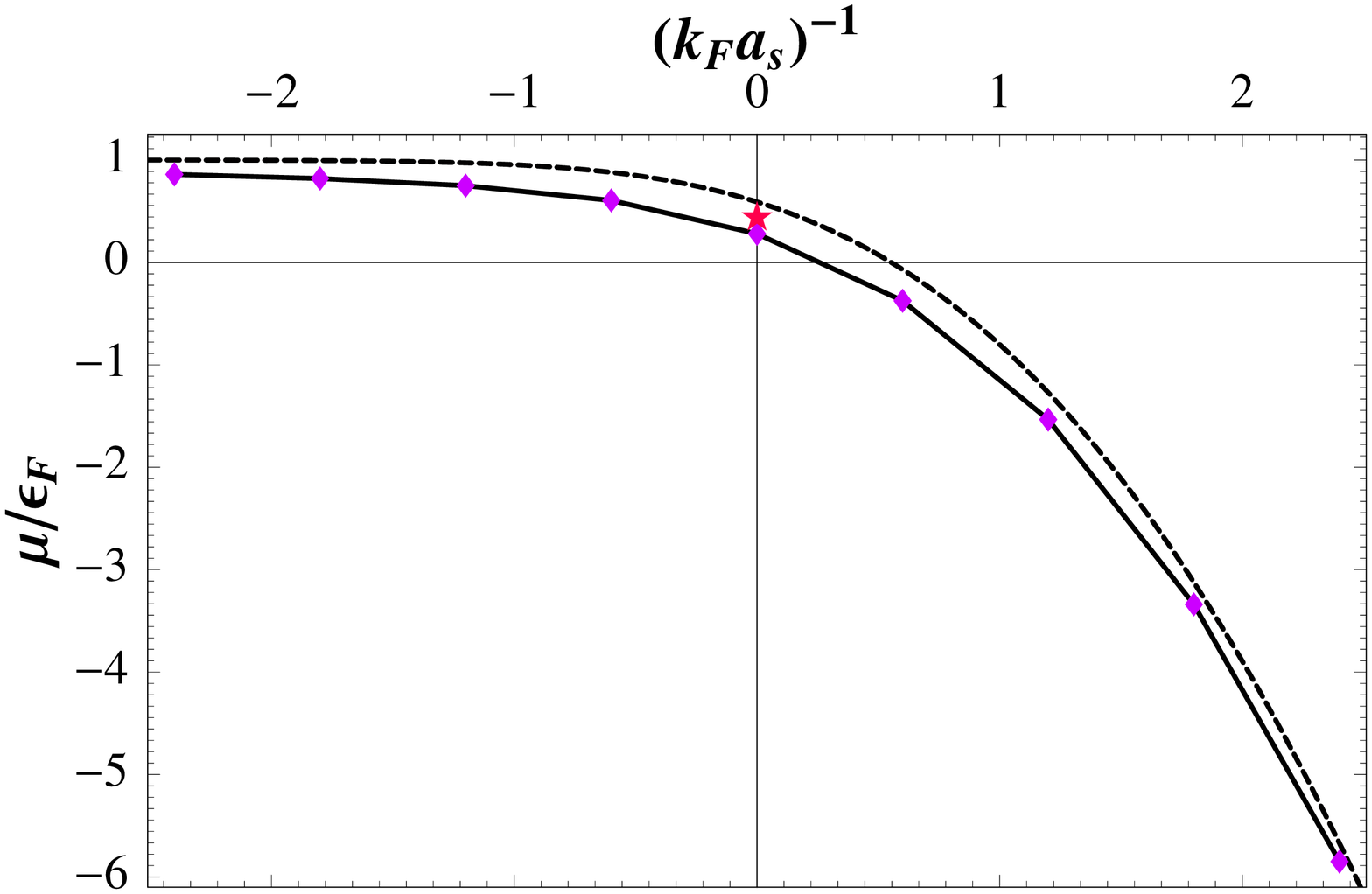}}
\put(0,0){\includegraphics[width=8.5cm]{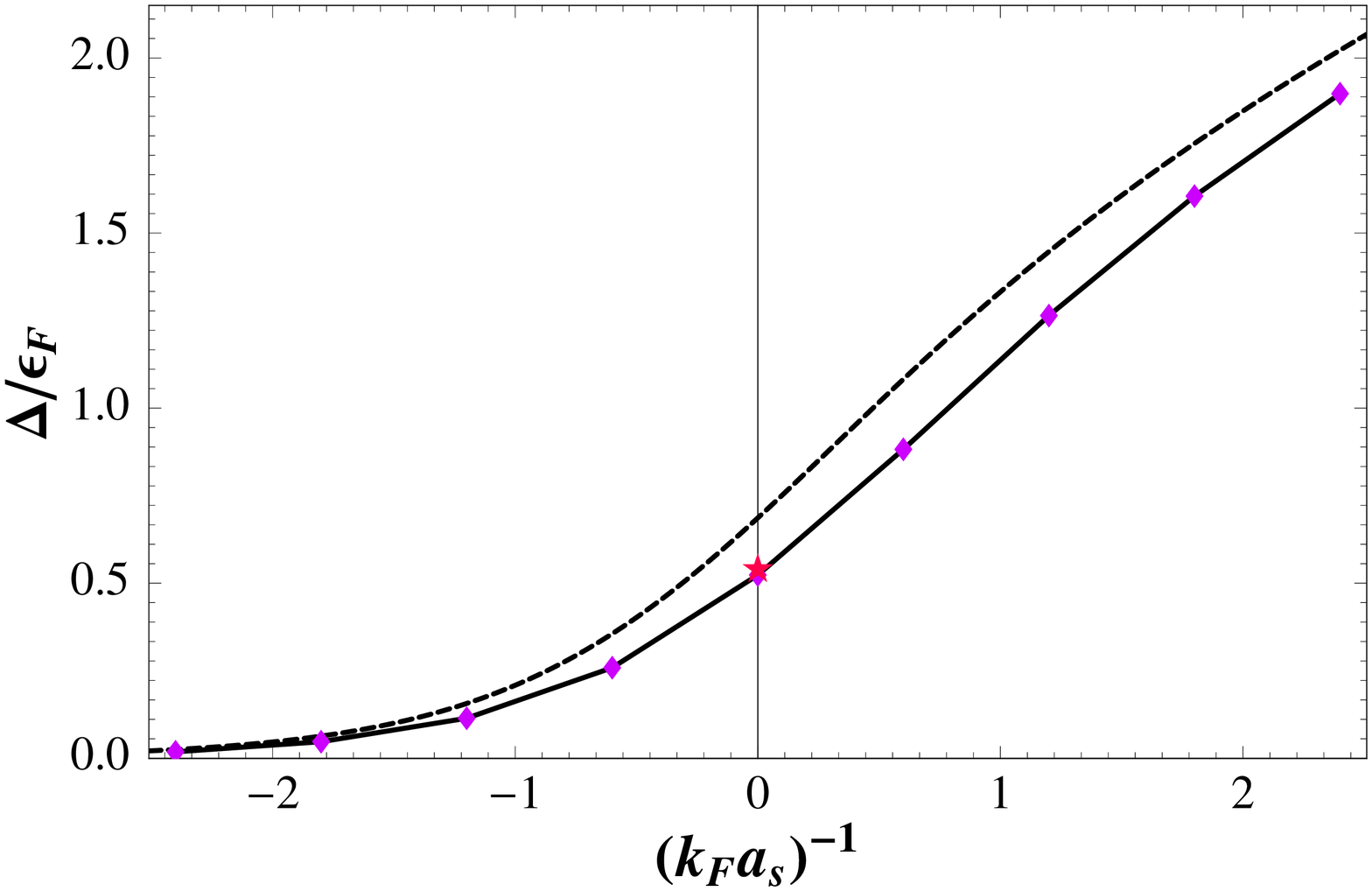}}
\end{picture}
\caption{Chemical potential $\mu$ and order parameter $\Delta$ as a
function
  of $(k_F a_s)^{-1}$. The dashed lines are the $\NS \to \infty$
 results for $\mu$ and $\Delta$. The diamond symbols include the
$\curO(1/\NS)$ corrections, evaluated at $\NS=1$.
 The solid lines are a guide to the eye. The star symbols at unitarity are the results of quantum Monte Carlo calculation from Ref. \onlinecite{Carlson2003}.}
\label{Figure1}
\end{center}
\end{figure}

Our results are predictions for the lowest nontrivial order in $1/\NS$
for an array of physical quantities at and around the unitary point.
At unitarity $\as^{-1} = 0$, a Fermi gas at density $n$ per fermionic
flavor is characterized by a single energy scale $\eF=( 3 \pi^2
n)^{2/3}/(2m)$.  In this regime, we find that, in the symmetry-broken
(paired-superfluid) phase at $T=0$, the chemical potential $\mu$,
the order parameter $\Delta$,
and the excitation gap $\Delta_{\rm exc}$
 to order $1/\NS$ are given by
\begin{align}
\label{1.2a}
\Delta/\eF                            &=0.6864 \numba/\NS + \curO(1/\NS^2), 
\\
\Delta_{\rm exc}/\eF                   &= 0.6864 -0.196/\NS + \curO(1/\NS^2), 
\\
\mu/\eF                               &=0.5906 \numbb/\NS +
\curO(1/\NS^2), \label{1.2b}
\end{align}
with the first $\NS\to\infty$ term corresponding to the well-known
mean-field theory result \cite{Eagles1969}. 
As we discuss in the Appendix, although $\Delta = \Delta_{\rm exc}$ in the $N=\infty$ limit,
at subleading order these quantities  differ, as seen in Eq.~(\ref{1.2a}).

By using scaling
arguments, it can be shown that, at unitarity, the ratio $\mu/\eF$ is
equal to $\xi$, defined by Eq. (\ref{1.1}). Substituting $\NS=1$ in
Eqs. (\ref{1.2a}) and
(\ref{1.2b}) gives our predictions %
\begin{align}\label{1.2aa}
\Delta/\eF                            &=0.523, \\
\Delta_{\rm exc}/\eF                            &=0.490, \\
\xi                               &=0.279 . \label{1.2ba}
\end{align}
 We have also computed the critical
temperature $T_c$ and the chemical potential $\mu$ at $T_c$ to be
given by
\begin{align}
\label{1.3a}
k_B T_c/\eF                          &=0.4964 \numbc/\NS + \curO(1/\NS^2), \\
\mu/\eF                              &=0.7469 \numbd/\NS +
\curO(1/\NS^2). \label{1.3b}
\end{align}

The $1/\NS$-expansion can also be applied to the problem of a
unitary Fermi gas with population imbalance $\delta
n=n_\uparrow-n_\downarrow$ between the densities $n_\uparrow$,
$n_\downarrow$ of the two hyperfine fermion components, that has
been realized in experiments \cite{Zwierlein2006,Partridge2006,Shin}
and has been a focus of intense theoretical activity; see e.g.,
Ref.\onlinecite{SR2006} and references therein.
Mean-field theory predicts~\cite{Bedaque,Caldas2004,SR2006,Parish2006}, that,
for arbitrarily small polarization $P \equiv \delta n/n$, a Fermi
gas in the unitary and positive-detuning regimes phase separates
into a polarized normal (unpaired) phase and an unpolarized
paired-superfluid phase, with the average density and polarization
equal to the experimentally imposed values.  Such phase separation
has, in fact, already been observed experimentally.  We find that,
within the $1/\NS$-expansion (consistent with mean-field theory) at
unitarity, the paired superfluid is unstable to phase separation for
an arbitrarily small polarization, i.e., $P_{c1} = 0$. \cite{SR2006}

For sufficiently large $P$, or, equivalently, sufficiently large
applied \lq\lq Zeeman field\rq\rq\ $h = \frac{1}{2}(\mu_\uparrow -
\mu_\downarrow)$ a uniform polarized normal phase is stable and the
corresponding critical values $P_{c2}$ and $h_{c2}$ are also
universal quantities at unitarity.  We find their $1/\NS$-expansion
to be given by
\begin{align}
\label{1.4a}
h_{c2}/\eF                                &=0.6929 \numbe/\NS + \curO(1/\NS^2),\\
\label{1.4b}
P_{c2}                                      &=0.9326 \numbf/\NS + \curO(1/\NS^2),\\
\mu/\eF                                &=0.8585 \numbg/\NS +
\curO(1/\NS^2), \label{1.4c}
\end{align}
where the chemical potential is evaluated at $h_{c2}$.

This paper is organized as follows. In Sec. \ref{SECTION2}, we
introduce the one-channel model and its generalization to the $\NS$
flavor case. In Sec. \ref{SECTION3}, we show that the solution for
$\NS \to \infty$ is equivalent to the mean-field solution of
the $\NS=1$ model. We then take into account, systematically, the
fluctuation corrections to  leading order in $1/\NS$ to the
internal energy, excitation gap, the critical temperature and the
upper-critical polarization, and extend our results to the vicinity
of the unitary point. In Sec. \ref{SECTION4}, we discuss our results
and compare them to recent measurements and predictions of other
theoretical approaches.  In Sec.  \ref{SECTION5}, we provide a brief
summary.

\section{Model}
\label{SECTION2}

We study a three-dimensional resonant Fermi gas confined to a box of
size $L \times L \times L=V$, described by a
Hamiltonian\cite{GRreview} (setting $\hbar=1$)
\begin{align}
\ham= &\sum_{\s=\uparrow,\downarrow}
\int d^3 \r \; \psid_\s (\r) \left( -\frac{\nabla^2_{\r}}{2 m} - \mu_\sigma
\right) \psipd_\s (\r) \notag \\
& +  \lambda \int d^3 \r  \; \psid_\uu(\r) \psid_\dd
(\r) \psipd_\dd(\r) \psipd_\uu(\r), \label{m1.1}
\end{align}
where $\lambda<0$ is the fermionic interaction and $\psid_\s(\r),
\psipd_\s(\r)$ are, respectively, the fermion creation and
annihilation operators at position $\r$ and hyperfine state
(\lq\lq spin \rq\rq -1/2) $\s$, which obey the usual anticommutation relation
$\left\{ \psipd_\s (\r), \psid_\si (\rp) \right\} = \delta(\r-\rp)
\delta_{\s,\si}$.  The chemical potential $\mu_\s$ fixes the average
density $n_\s$ of particles with spin $\s$.  For simplicity we now set
$\mu_\uparrow = \mu_\downarrow = \mu$, although later we shall allow
the possibility of a nonzero chemical potential difference in order to
study a polarized Fermi gas.

The partition function $Z= {\rm Tr} \left[ e^{-\beta \ham} \right]$
gives the free-energy density
\be \FFF= -\frac{1}{\beta V} \log Z, \label{m1.2} \ee
where $\beta=1/T$ is the inverse temperature, taking $k_B=1$
throughout.

The two-particle scattering length $\as$ is related to the strength of
the atomic interaction $\lambda$ via the relation
\be \frac{m}{4 \pi
  \as}= \frac{1}{\lambda} + \frac{1}{V} \sum_\k  \frac{1}{2
  \eps_\k},
\label{m1.3}
\ee
where $\eps_\k= k^2/(2m)$ is the free fermion dispersion (kinetic
energy), taking $\hbar=1$ throughout. In the second term on the
right side of Eq. (\ref{m1.3}) an ultraviolet cutoff $\Lambda \sim
1/r_o $ is implied, set by the effective interaction range $r_o \ll
1/k_F$~\cite{narrownote,GRreview}.

We now proceed with the generalization of Eq.~(\ref{m1.1}) to a model
of $\NS$ fermion flavors:
\begin{align}
\ham= &\sum_{i=1}^{\NS}\sum_{\s=\uparrow,\downarrow}
\int d^3 \r \; \psid_{i\s} (\r) \left( -\frac{\nabla^2_{\r}}{2 m} - \mu
\right) \psipd_{i\s} (\r) \notag \\
& +  \frac{\lambda}{\NS}\sum_{i,j=1}^{\NS} \int d^3 \r  \; \psid_{i\uu}(\r) \psid_{i\dd}
(\r) \psipd_{j\dd}(\r) \psipd_{j\uu}(\r),
\label{m1.1prime}
\end{align}
that is clearly equal to Eq.~(\ref{m1.1}) for the case of $\NS = 1$
and possessing invariance under the symplectic group $Sp(2N)$ (see
Ref.~\onlinecite{symplectic}).
As discussed in Sec.~\ref{Motivation}, the benefit of this expansion
to $\NS$ fermion flavors is that an exact solution may be obtained
in the large-$\NS$ limit.  To see this, we write the partition
function in terms of an imaginary-time coherent-state functional
integral
\be Z=\int  D \psio_{i\s}(x) D \psipo_{i\s}(x) \exp(-\SNS).
\label{m1.5}
\ee
with $\psio_{i\s}(x),\psipo_{i\s}(x)$ Grassmann fields labeling the
corresponding coherent state at a space-time point $x=(\r,\tau)$,
and with the imaginary-time action $\SNS$ given by
\begin{align}
  \SNS &= \sum_{i=1}^{\NS} \sum_{\s=\uparrow,\downarrow}
\int_0^\beta d \tau \int d^3 \r
  \; \psio_{i\s} (x)\left(
    \partial_\tau -\frac{\nabla^2_{\r}}{2 m} - \mu \right) \psipo_{i\s}(x) \notag \\
  &+ \frac{\lambda}{\NS} \sum_{i,j=1}^{\NS} \int_0^\beta d \tau \int
  d^3 \r \; \psio_{i\uu}(x) \psio_{i\dd}(x) \psipo_{j\dd}(x)
  \psipo_{j\uu}(x).
\label{m1.4}
\end{align}

We decouple the atomic interaction via a standard Hubbard-Stratonovich
transformation, that relies on the Gaussian integral
\begin{align}
  e^{-\frac{\lambda}{\NS} \phio(x) \phipo(x)}=
  \frac{-\NS}{\pi \lambda} \int & d b(x) d b^\ast(x)
\exp\big(\frac{\NS}{\lambda}b^\ast(x) b(x) \notag  \\
  & + b(x) \phio(x) + b^\ast(x) \phipo(x) \big),
\label{m1.6}
\end{align}
which we shall utilize with $\phi(x)$ given by a superposition of
$\NS$ flavors of bilinear fermionic fields
\be
\phi(x)= \sum^{{\NS}}_{i=1} \psi_{i\dd}(x) \psi_{i\uu}(x).
\ee
With this transformation, the fermion fields appear quadratically and
can therefore be formally integrated out. The partition function is
then given by a functional integral over the boson field $b(x)$,
\be Z = Z^{-1}_b \int D b(x) D b^\ast(x)e^{-S[b]},
\label{m1.9}
\ee
with the effective bosonic action given by
\bea &&S[b]=
-\NS\int_0^\beta d\tau \int d^3 \r \frac{b^\ast(x)b(x)}{\lambda}
\nonumber \\
&&\qquad \qquad
-\NS{\rm Tr} \log\big(-
G^{-1}\left[b(x),b^\ast(x) \right]\big) \label{m1.7},
\eea
where the trace is over space, imaginary time, hyperfine and flavor
states, the Green's function (written in Nambu representation for
the fermions) is
\be
G^{-1}(x)=
\begin{pmatrix}
-\partial_\tau + \frac{\nabla^2}{2m} + \mu & b(x) \\
b^\ast(x)& -\partial_\tau - \frac{\nabla^2}{2m} - \mu
\end{pmatrix},
\label{m1.8}
\ee
and the Hubbard-Stratonovich normalization factor $Z_b=\int D b(x)
D b^\ast(x) \exp \left(\NS\int_0^\beta d \tau \int d^3 \r b^\ast(x)
  b(x)/\lambda \right)$. Note that in the limit where $\NS=1$ the
effective action reduces to the one-channel model derived by S\'a de
Melo, \textit{ et al.} \cite{Sademelo1993}.

We use this formulation to compute a number of thermodynamic
quantities that characterize the resonant unitary Fermi gas. To this
end, we observe that in the limit ${\NS}\to \infty$, these
expectations values are dominated by the saddle point of the
functional integral.  This is the origin of the aforementioned claim
of an exact solution in the limit of large-$\NS$.  As is well-known,
such a saddle-point evaluation amounts to the standard BEC-BCS
mean-field approximation to the one-channel model~\cite{Leggett1980}
that is only {\it approximately\/} valid for the case $\NS = 1$.  At
large $\NS$ , fluctuations around this saddle point are small in the
parameter $1/\NS$, generating $\curO (1/\NS)$ corrections, that
organize into a systematic expansion in powers of $1/\NS$ for the
partition function and physical quantities derived from it.

By representing the bosonic Green's function by a line, the
partition function is given by a series of closed loops. The higher
order terms in $1/{\NS}$ can be classified according to the number
of loops for a particular diagram.  Thus the $1/{\NS}$-expansion is
equivalent to a so-called loop expansion.

The implementation of the loop expansion is performed around the
(possibly complex) saddle point $\bo $, where $\bo $ is the
mean-field value of $b(x)$. At low temperature, that is our focus
here, the saddle point yields a non-zero value for $\bo $. The
complex field fluctuations $\b (x)$ around the saddle point are
defined as \be \frac{1}{\sqrt{{\NS}}} \b(x) \equiv b(x) - \bo
\label{m1.10}. \ee The Green's function matrix can be formally
separated into the saddle point and fluctuation contributions, \be
G^{-1}(x)=G^{-1}_{(0)}(x) +\frac{1}{\sqrt{{\NS}}} G^{-1}_{(1)}(x)
\label{m1.11}, \ee where \be G^{-1}_{(0)} (x)=
\begin{pmatrix}
-\partial_\tau + \frac{\nabla^2}{2m} + \mu & \bo  \\
\bo^{\ast}& -\partial_\tau - \frac{\nabla^2}{2m} - \mu
\end{pmatrix}
\label{m1.11a},
\ee
and
\be
G^{-1}_{(1)} (x)=
\begin{pmatrix}
0 & \b (x) \\
\b^{\ast}(x)& 0
\end{pmatrix}
\label{m1.11b},
\ee
The bosonic action $S[b]$ can be expanded in powers of $1/{\NS}$
\begin{align}
\frac{S[b]}{\NS}=&
-  \beta V
\frac{|\bo |^2}{\lambda} - {\rm Tr} \left[ \log \left( -G_{(0)}^{-1} \right) \right] \notag \\
&\hspace{-.5cm}- \int_0^{\beta} d
\tau \int d^3 \r \;\left[  \frac{1}{\sqrt{{\NS}}}
\frac{ \bo  \b^\ast(x) +\bo  ^\ast \b(x)}{\lambda} +
\frac{1}{{\NS}} \frac{|\b(x)|^2}{\lambda} \right] \notag \\
&\hspace{-.5cm} +  \sum_{m=1}^{\infty} \frac{(-1)^m}{m} \frac{1}{{\NS}^{m/2}} {\rm
    Tr} \left[ \left( G_{(0)} G^{-1}_{(1)}[\b(x)] \right)^m \right]
\label{m1.13}.
\end{align}

Up to this point, all the transformations of the partition function
are exact. An approximation is made when only a finite number of terms
in the action are considered, and the small parameter $1/\NS$ provides
a systematic way to organize this expansion.  To analyze the problem
to lowest nontrivial order in $1/\NS$, it is sufficient to include
terms up to $m=2$ in the action, yielding an approximate bosonic action
\be \frac{S[b]}{\NS}= \SOO + \frac{1}{\NS} \STT + \dots
\label{m1.14}, \ee where
\be \SOO=-\beta V \frac{|\bo |^2}{\lambda}-{\rm Tr}\left[ \log
\left(-
    G^{-1}_{(0)} \right) \right]
\label{m1.15},
\ee
and
\begin{align}
\STT =& - \int_0^\beta d \tau \int d^3 \r
\frac{|\b(x)|^2}{\lambda} \notag \\
& +\frac{1}{2} {\rm Tr}
\left[ G^{-1}_{(1)}[\b(x)] G_{(0)} G^{-1}_{(1)}[\b(x)] G_{(0)} \right]
\label{m1.16},
\end{align}
that we shall analyze in the remainder of the manuscript. We note
that, as usual, terms linear in $\b$ (the $m=1$ terms) automatically
vanish by a virtue of the definition of the saddle-point as the
extremum of the action $\SOO$.

\section{large-$\NS$  expansion}
\label{SECTION3} In the present section, we use the results of
Sec.~\ref{SECTION2} to construct an expansion of the free-energy
density $\FFF$ to leading order in small $1/\NS$ of the form
\be \FFF  =N \FFO+ \FFT+ \dots \label{free}, \ee
where the leading-order term $\FFO$ comes from a saddle-point
approximation to the bosonic functional-integral expression, Eq.~(\ref{m1.9}),
for the partition function.

\subsection{Saddle-point approximation: Mean-field level}
\label{SEC:saddle} In this section, we compute the lowest order
approximation to the free-energy density, namely the first term of
Eq.~(\ref{free}), that is exact in the $\NS \to \infty$
limit. This corresponds to an evaluation of the free-energy density
within the saddle-point approximation $\FFO=-\frac{1}{\beta V} \log
Z^{(0)}=\frac{1}{\beta V}\SOO$, which gives
\be \FFO= -\frac{|\bo|^2}{\lambda}-\frac{1}{V} \sum_\k \left( E_\k -
\xi_\k \right)- \frac{2}{\beta V} \sum_{\k} \log \left[ 1+ e^{-\beta
E_{\k}} \right] \label{m2.1}, \ee
where we included an overall constant in the second term so that for
$\Delta \to 0$, $\FFO$ reduces to the free-energy density of a free
Fermi gas.  Here, $E_{\k}=\sqrt{\xi^2_{\k} +|\bo |^2}$ is the
spectrum of the quasiparticles and $\xi_\k=\epsilon_\k-\mu$.  Notice
that this $\NS \to \infty$ limit yields a result for the
free energy (and therefore all subsequent thermodynamics quantities)
that is identical to the usual mean-field treatment based on a BCS
ansatz for the ground-state wavefunction.

The gap equation, determined by minimizing the free energy with
respect to the gap, $\frac{ \partial f}{ \partial \bo}=0$, yields
\be
-\frac{m \bo}{4 \pi \as} = \frac{\bo}{V} \sum_\k \left(
\frac{\tanh
 (\beta  E_\k /2 )}{2 E_{\k}} -\frac{1}{2 \epsilon_\k} \right)
\label{m2.2}.
\ee
The total atom density is $n \times \NS$, where $n$ is the atom
density of each of the fermion flavors. It is determined through a
Lagrange multiplier via $n \NS = -\frac{ \partial f}{ \partial \mu}$,
which yields for $\NS \to \infty$
\be n= \frac{1}{V} \sum_\k \left( 1- \frac{\xi_k}{E_\k} \tanh(\beta
E_\k/2) \right) \label{m2.3}.
 \ee
This number equation, along with Eq.~(\ref{m2.2}), describes a
resonant Fermi gas within the mean-field approximation.

The solution to these equations at zero and finite temperature has
been discussed at length in, e.g., Refs.
\onlinecite{Leggett1980,Nozieres1985,Sademelo1993,Marini1997,GRreview}.
For completeness, below we review this mean-field solution near
unitarity at zero temperature, near the transition temperature $T_c$,
and its generalization to an imposed polarization \cite{SR2006}.

\subsubsection{Zero-temperature limit}
\label{SEC:ZTL}
At zero temperature, the mean-field ground state at unitarity, $(k_F
\as)^{-1}=0$, is a superfluid characterized by the BCS wavefunction,
with the gap and chemical potential given by
\begin{align}
\Delta\qao/\eF       &=0.6864, \\
\mu\qao/\eF          &=0.5906 \label{m2.5},
\end{align}
simply obtained by solving Eqs.~(\ref{m2.2}) and ~(\ref{m2.3}) at
$T=0$.  Here, the superscript and subscript refer to the
$1/\NS = 0$ and $T=0$ limits, respectively.

\subsubsection{Near $T_c$}
\label{SEC:neartc}
At finite temperatures, a second-order superfluid-to-normal transition
takes place at a critical temperature $T_c$, that in the
$\NS \to\infty$ limit is given by
\begin{align}
k_B T\qat/\eF      & = 0.4965  , \\
\mu\qat/\eF        & = 0.7469 , \label{m2.4}
\end{align}
obtained by solving these equations at $\Delta =0^{+}$; $\mu\qat$ is
the chemical potential at the transition as determined by the
fixed-density condition Eq.~(\ref{m2.3}).

\subsubsection{Polarized Fermi gas}
\label{SEC:pfg}
We can also study a polarized Fermi gas (one with
population imbalance between the spin components) in the
saddle-point approximation, achieved through coupling the fermions
to an applied \lq\lq Zeeman\rq\rq\ field.
Physically, this corresponds to imposing different chemical
potentials, $\mu_{\uu / \dd }=\mu \pm h $ for the up and down spins,
and yields the free-energy density
\begin{align}
\FFO=& -\frac{|\bo|^2}{\lambda}-\frac{1}{V} \sum_\k \left( E_\k -
\xi_\k \right) \notag \\
& -  \frac{1}{\beta V} \sum_{\sigma=\pm} \sum_{\k} \log \left[ 1+
  e^{-\beta (E_{\k}+\sigma h)} \right]
\label{m2.6}.
\end{align}
The detailed mean-field phase diagram found by minimizing $\FFO$ has
been discussed in detail elsewhere~\cite{SR2006,Parish2006}. The
relevant parameters characterizing the phase diagrams are the detuning
and $h$, or equivalently (and more directly related to experiments at
fixed species imbalance) as a function of an imposed polarization
\be
P = \frac{m}{n} = \frac{\frac{\partial f}{\partial h}}{\frac{\partial f}{\partial \mu}},
\label{eq:polarization}
\ee
with the density imbalance $m$ and density $n$.
We are particularly interested in the region near unitarity, where for
sufficiently large $h$ at fixed $\mu$ there is a first-order phase
transition between the superfluid ($s$) and normal ($n$) states.
At fixed imposed density $n$, the first-order transition opens up into
a regime $h_{c1}<h<h_{c2}$ of phase separation, where on the BCS side
and around the unitary point a gapped unpolarized superfluid and a
polarized Fermi gas coexist.  For $h<h_{c1}$, the system is in the
homogeneous fully-paired superfluid state. At unitarity, the
lower-critical polarization for entering the regime of phase
separation is $P_{c1} =0$ ~\cite{SR2006}.  For $h>h_{c2}$, the normal
state is stable. The instability to phase-separation from the normal
state takes place at $h=h_{c2}$ when the free energies of the normal
($\NNN$) and superfluid ($\SSS$) states cross, subject to the
constraint on $\mu$ that the normal-state number equation is
satisfied.  Thus, we equate $\FFS=\FFN$, using the $\NS \to \infty$
approximation to the superfluid-state free-energy density
$\FFS\approx\FFO(\Delta)$ (with $\Delta$ determined by the
minimization $\partial \FFS/\partial \Delta=0$), and the normal state
free-energy density given by $\FFN=\FFS(0)$.  Solving these equations
at zero temperature, we find $\mu$, $h$ and $\Delta$ at the transition
(in agreement with earlier work~\cite{SR2006,Parish2006})
\begin{align}
\nonumber
\mu\qah/\eF          &=0.8586, \\
\nonumber
h\qah/\eF            &=0.6930, \\
\Delta\qah/\eF       &=0.9979 \label{m2.7}.
\end{align}
The upper-critical polarization is given by combining the
normal-state polarization Eq.~(\ref{eq:polarization}) with $\FFF$
given by $\FFN$ (taking $T=0$)
\be
P=\frac{(1+h/\mu)^{3/2}-(1-h/\mu)^{3/2}}{(1+h/\mu)^{3/2}+(1-h/\mu)^{3/2}},
\label{m2.9}
\ee
with the location of the transition defined by  Eq.~(\ref{m2.7}), yielding
 the mean-field result
\begin{equation}
P_{c2}=0.93261,
\end{equation}
at unitarity.

The quantities that we have computed in this section within the $\NS
\to \infty$ limit (that is equivalent to the BCS mean-field theory),
are {\em universal}, i.e., independent of the microscopic
interactions and can be obtained as derivatives of a universal free
energy that is given by the Fermi energy (set by the atomic density
$n$) times a system-independent scaling function of dimensionless
variables such as $T/\eF$ and $h/\eF$.  For example, taking $h = 0$
for simplicity, based on dimensional grounds, the free-energy
density at unitarity can only depend on $T$ and the density $n$
(which we represent through $\eF$, see Eq.(\ref{1.1}) )
\be f(T,n)=\frac{3}{5}
\frac{(3\pi^2)^{2/3}}{2 m} n^{5/3} g \left(\frac{k_B T}{(3\pi^2)
    n^{2/3}/2m} \right),
\ee
where $g(x)$ is a dimensionless function.  At zero temperature, the
internal energy per particle [quoted in Eq.~(\ref{1.1})]
$\MEANE=f(T=0)/n$ is $\frac{3}{5}g(0) \eF$, while the chemical
potential $\mu=\frac{\partial f}{\partial n}$ is equal to $\mu= g(0)
\eF$.  This leads to $\MEANE=\frac{3}{5} \mu$, that together with
Eq.~(\ref{m2.5}) gives our lowest order $\NS \to\infty$
(mean-field) estimate of the parameter $\xi$:
\be
\xi =g(0) \approx 0.5906.
\ee
Other quantities that are connected to $\xi$ at $T=0$ are the bulk
modulus $B\equiv n^2\partial^2 f/\partial n^2$ given by:
\begin{eqnarray}
\label{B}
B&=&\frac{2}{3} \xi n \eF,
\end{eqnarray}
and the first- (isothermal) sound $v=\sqrt{B/n m}$.

In a similar fashion, one can derive the equation of state of the
system and determine relations between the entropy, pressure and
internal energy~\cite{Burovski2006a}. Since these relations follow
from scaling arguments, they hold for any number of fermion flavors.
Consequently, these relations are expected to be preserved order
by order in the $1/\NS$-expansion.

Having obtained leading-order results in $1/\NS$ (already reported
in the literature as they correspond to the standard mean-field
theory, exact for the $\NS=\infty$ model), in the next subsection we
proceed to calculate the leading-order corrections in small $1/\NS$.
We shall see that the large-$\NS$ expansion of the one-channel model
organizes subleading corrections to mean-field theory, essentially
amounting to the random-phase approximation.

\subsection{Leading order in $1/{\NS}$: Random Phase Approximation (RPA)}

We now consider the leading-order corrections in $1/N$ to the results
of the previous section, which requires an evaluation of the
subleading contribution to the free energy.  This is given by a
Gaussian integration over $\b$ fluctuations around the saddle point,
governed by the effective action Eq.~(\ref{m1.16}), explicitly given by
\be \STT=\frac{1}{2} \sum_q
\begin{pmatrix}
 \b^\ast(q) & \b(-q)
\end{pmatrix}
\begin{pmatrix}
A(q) & B(q)\\
B^\ast(q) & A(-q)
\end{pmatrix}
\begin{pmatrix}
\b(q)\\
\b^\ast(-q)
\end{pmatrix}
\label{m3.3},
\ee
where $q=(\q,\Omega_{\ell})$ and the bosonic Matsubara frequency
$\Omega_{\ell}= 2 \pi \ell/\beta$ with $\ell$ an integer.  The matrix
elements of the polarization matrix are
\begin{align}\label{m3.4a}
A(q)&= -\frac{1}{\lambda}+\frac{1}{\beta V} \sum_k
\left(G_{(0)}(k+q)\right)_{11} \left( G_{(0)}(k)\right)_{22}, \\
B(q)&= \frac{1} {\beta V}  \sum_k  \left( G_{(0)}(q+k)\right)_{21}
\left( G_{(0)}(k)\right)_{21} \label{m3.4b},
\end{align}
which satisfy the relations $A(q)= A^\ast(-q)$, $B(q)=B(-q)$.

The quantity $G_{(0)}$ appearing in Eqs.~(\ref{m3.4a}) and
~(\ref{m3.4b}) is the standard saddle-point approximation to the
single-particle Green's function in the BCS state
\begin{align}
G_{(0)}(k)&=\frac{-1}{\omega_{\ell}^2+E^2_\k}
\begin{pmatrix}
i \omega_{\ell} + \xi_\k  & -\bo \\
-\bo^{\ast}    & i \omega_{\ell} - \xi_\k
\end{pmatrix} \notag \\
&=
\begin{pmatrix}
\frac{u^2_\k}{i \omega_{\ell}-E_{\k}} +\frac{v^2_\k}{i \omega_{\ell}+E_{\k}} & -\frac{u_\k v_\k}{i \omega_{\ell}-E_{\k}} +\frac{u_\k v_\k}{i \omega_{\ell}+E_{\k}} \\
-\frac{u_\k v_\k}{i \omega_{\ell}-E_{\k}} +\frac{u_\k v_\k}{i
\omega_{\ell}+E_{\k}}  & \frac{v^2_\k}{i \omega_{\ell}-E_{\k}}
+\frac{u_\k^2}{i \omega_{\ell}+E_{\k}}
\end{pmatrix},
\end{align}
where
\begin{align}
u_\k&=\sqrt{\frac{1}{2}\left(1+\frac{\xi_\k}{E_\k} \right)}, \\
v_\k&=\sqrt{\frac{1}{2}\left(1-\frac{\xi_\k}{E_\k} \right)}
\label{m3.5},
\end{align}
are the usual BCS coherence factors and $k=(\k,i\omega_{\ell})$ and
$\omega_\ell=(2\ell+1)\pi/\beta$ are the fermionic Matsubara
frequencies and for simplicity we have taken $\Delta$ to be real.
Evaluation of these normal and anomalous particle-particle bubbles
yields, respectively,
\begin{align} \label{m3.6}
      A(q) & =-\frac{1}{\lambda}
          +\frac{1}{2 V} \sum_{\bf k} \\
        &\Big[
        \left(\tanh (\beta E_\k/2) +\tanh(\beta E_{\k+\q}/2) \right) \notag \\
        &\left( -
        {
        u^2_{\k+\q} u_{\k}^2
        \over
        E_{\k+\q} + E_{\k} - i \Omega_{\ell}
        }
        -
        {
        v^2_{{\bf k}+{\bf q}} v_{{\bf k}}^2
        \over
        E_{{\bf k}+{\bf q}} + E_{{\bf k}} + i \Omega_{\ell}
        }
        \right) \notag \\
        & +
        \left(\tanh (\beta E_\k/2) -\tanh(\beta E_{\k+\q}/2) \right) \notag \\
        &
        \left(
        {
        u^2_{\k+\q} v_{\k}^2
        \over
        E_{\k} - E_{\k+\q} + i \Omega_{\ell}
        }
        +
        {
        v^2_{{\bf k}+{\bf q}} u_{{\bf k}}^2
        \over
        E_{{\bf k}} - E_{{\bf k}+\q} - i \Omega_{\ell}
        }
        \right) \Big]  \notag
       ,
\end{align}
\begin{align} \label{m3.7}
      B(q) & =
         \frac{1}{2 V} \sum_{\bf k}  u_{\k+\q} u_{\k}
         v_{\k+\q} v_{\k}\\
        &\Big[
        \left(\tanh (\beta E_\k/2) +\tanh(\beta E_{\k+\q}/2) \right)\notag \\
       & \left(
        {
        1
        \over
        E_{\k+\q} + E_{\k} + i \Omega_{\ell}
        }
        +
        {
        1
        \over
        E_{{\bf k}+{\bf q}} + E_{{\bf k}} - i \Omega_{\ell}
        }
        \right) \notag \\
        & +
        \left(\tanh (\beta E_\k/2) -\tanh(\beta E_{\k+\q}/2) \right) \notag \\
        & \left(
        {
        -1
        \over
        -E_{\k} + E_{\k+\q} - i \Omega_{\ell}
        }
        +
        {
        -1
        \over
        -E_{{\bf k}} + E_{{\bf k}+\q} + i \Omega_{\ell}
        }
        \right) \Big]
       . \notag
\end{align}

In terms of these the Gaussian functional integral over the $\b(x)$
and $\b^\ast(x)$ fields finally gives the $1/\NS$ contribution to the
free-energy density
\be \FFT = \frac{1}{2 \beta V} \sum_{q} e^{i \Omega_{\ell} \delta}
\log \left[ \lambda^2 \left( |A(q)|^2-|B(q)|^2 \right) \right]
\label{m3.8}, \ee
where the function is evaluated at an imaginary time $\delta=0^+$.

When inserted into Eqs.~(\ref{free}), and using Eq.~(\ref{m2.1}),
Eq.~(\ref{m3.8}) yields an explicit expression for the free-energy
density of a resonantly-interacting Fermi gas to leading order in
$1/\NS$.  We now proceed to compute the effect of the $1/\NS$
corrections in various limiting regimes, starting with $T=0$.

\subsubsection{Zero-Temperature Limit}

At zero temperature, the energy density (to subleading order) is
\be \frac{\EEE}{{\NS}} =\EEO+\frac{1}{{\NS}} \EET+ \dots
\label{m3.10}, \ee
 where $\EEO=\FFO(T=0)$ and
\be \EET= \frac{1}{2 V} \sum_{\q} \int_{-\infty}^{\infty} \frac{d
\Omega}{2\pi}  \; e^{i \Omega \delta} \log \left[ \lambda^2
\left(|A(\q,i\Omega)|^2- |B(\q,i\Omega)|^2 \right) \right]
\label{m3.11}. \ee
At large $\NS$, the chemical potential $\mu$ and gap parameter
$\Delta$ are very close to their $\NS \to \infty$ values,
within a correction of order $\curO(1/\NS)$ :
\begin{align}
\label{m3.12a}
\Delta&= \Delta\qao+ \frac{1}{\NS}\delta \Delta + \dots, \\
\mu&= \mu\qao+ \frac{1}{\NS} \delta \mu +\dots \label{m3.12b}.
\end{align}
Inserting Eqs.~(\ref{m3.12a}) and (\ref{m3.12b}) into
Eq.~(\ref{m3.11}) and using the gap $0 = \partial \EEE/\partial
\Delta$ and number $n \NS  = - \partial \EEE/\partial \mu$
equations, we find that $\delta \Delta$ and $\delta \mu$ satisfy
\be
\begin{pmatrix}
\partial_{\mu \mu} \EEO & \partial_{\mu \Delta} \EEO \\
\partial_{\Delta \mu} \EEO & \partial_{\Delta \Delta} \EEO
\end{pmatrix}
\begin{pmatrix}
\delta \mu \\
\delta \Delta
\end{pmatrix}
=-
\begin{pmatrix}
\partial_\mu \EET \\
\partial_\Delta \EET
\end{pmatrix}
\label{m3.13},
\ee
where all the derivatives (indicated by the shorthand notation
$\partial_{a b} \equiv \frac{\partial}{\partial a}
\frac{\partial}{\partial b} $) are evaluated at the saddle point, i.e.
$\Delta=\Delta\qao$ and $\mu=\mu\qao$. Solving for $\delta \Delta$ and
$\delta \mu$ yields
\begin{align}
\begin{pmatrix}
\delta \mu \\
\delta \Delta
\end{pmatrix}=&\frac{-1}{ \partial_{\mu \mu} \EEO
\partial_{\Delta \Delta} \EEO - \partial_{\mu \Delta} \EEO \partial_{\Delta \mu} \EEO}
\notag \\
&\times \begin{pmatrix}
\partial_{\Delta \Delta} \EEO & -\partial_{\mu \Delta} \EEO \\
-\partial_{\Delta \mu} \EEO & \partial_{\mu \mu} \EEO
\end{pmatrix}
\begin{pmatrix}
\partial_\mu \EET \\
\partial_\Delta \EET
\end{pmatrix}
\label{m3.14},
\end{align}
thus expressing the shifts $\delta \mu$ and $\delta \Delta$ in the
chemical potential and gap in terms of various derivatives of the
integrals $\EEO$ and $\EET$.  At unitarity, these can be calculated
numerically and are given by
\begin{align}
 \partial_{\mu \mu} \EEO &= -1.0804 n/\eF   ,\\
\partial_{\mu \Delta} \EEO &= -1.2556  n/\eF  ,\\
\partial_{\Delta \Delta} \EEO &= 1.0804 n/\eF  , \\
\partial_{\mu} \EET &=  -0.542   n         ,       \\
\partial_{\Delta} \EET &= -0.216 n.
\label{m3.15}
\end{align}
Substituting Eq. (\ref{m3.14}) and the appropriate parameters into
Eqs. (\ref{m3.12a}) and (\ref{m3.12b}) gives the $1/\NS$ corrections to
$\mu$ and $\Delta$ quoted in Eqs.~(\ref{1.2a}) and (\ref{1.2b}) of the
Introduction.

\subsubsection{Near $T_c$}

Next, we consider the vicinity of the transition temperature $T_c$,
for which the saddle-point ($\NS \to \infty$) results were
presented in Sec.~\ref{SEC:neartc}.  The $1/\NS$ corrections to the
critical temperature can be evaluated by computing the Thouless
criterion ($\partial_\Delta f =0^{+}$) and particle number equation
from the free energy. Combining the two equations allows the
determination of the critical temperature and chemical potential
correction. Writing the saddle-point solution with the indices $\qat$,
we can parameterize the $1/\NS$ corrections to $T$ and $\mu$ near
$T_c$ via
\begin{align}
\label{m3.16a}
T&= T\qat+ \frac{1}{\NS}\delta T + \dots , \\
\mu&= \mu\qat+ \frac{1}{\NS} \delta \mu +\dots \label{m3.16b},
\end{align}
analogously to the zero-temperature case in the preceding subsection.
As in that case, the Thouless criterion and particle equation yield a
set of two coupled equations
\be
\begin{pmatrix}
\partial_{\mu \mu} \FFO & \partial_{\mu T} \FFO \\
\partial_{\Delta \mu} \FFO & \partial_{\Delta T} \FFO
\end{pmatrix}
\begin{pmatrix}
\delta \mu \\
\delta T
\end{pmatrix}
=-
\begin{pmatrix}
\partial_\mu \FFT \\
\partial_{\Delta} \FFT
\end{pmatrix}
\label{m3.17},
\ee
where all the derivatives are evaluated at $T=T\qat$ and
$\mu=\mu\qat$ and $\Delta=0^+$. Solving for $\delta T$ and $\delta
\mu$, we obtain
\begin{align}
\begin{pmatrix}
\delta \mu \\
\delta T
\end{pmatrix}=&\frac{-1}{ \partial_{\mu \mu} \FFO \partial_{\Delta T} \FFO - \partial_{\mu T} \FFO  \partial_{\Delta \mu} \FFO}
\notag \\
& \times
\begin{pmatrix}
\partial_{\Delta T} \FFO & -\partial_{\mu T} \FFO \\
-\partial_{\Delta \mu} \FFO & \partial_{\mu \mu} \FFO
\end{pmatrix}
\begin{pmatrix}
\partial_\mu \FFT \\
\partial_{\Delta} \FFT
\end{pmatrix}
\label{m3.18}.
\end{align}
At unitarity, we find the matrix elements of Eq.~(\ref{m3.18}) to be
\begin{align}
\FFO &=-0.678 \,n\eF,\\
\partial_{\mu \mu} \FFO &= -1.1720 n/\eF, \\
\partial_{\mu T} \FFO &=-1.2581 n/\eF,\\
\partial_{\Delta \mu } \FFO &=-1.5691 n\Delta/\eF^2 ,\\
\partial_{\Delta T} \FFO &= 2.3608 n\Delta/\eF^2 ,\\
\FFT & = -0.701   n\eF ,\\
\partial_\mu \FFT & =-2.339 n,\\
\partial_{\Delta} \FFT& = 2.195 n\Delta/\eF .
\label{m3.19}
\end{align}
Using Eq. (\ref{m3.18}) together with these matrix elements inside
Eqs.~(\ref{m3.16a}) and (\ref{m3.16b}) gives the $1/\NS$ correction to the
critical temperature and the chemical potential quoted in
Eqs.~(\ref{1.3a}) and (\ref{1.3b}) of the Introduction.

\subsubsection{Polarized Fermi gas}

We now determine the $1/\NS$ corrections, beyond the mean-field result
reviewed in Sec.~\ref{SEC:pfg}, to the upper critical polarization
$P_{c2}$.  Recall the two criteria to determine the upper-critical
polarization, above which the normal state is stable.  These are given
by equality of the superfluid and normal-state free energies (the
first-order transition condition) and the normal-state number
equation, that are, respectively, given by
\begin{align}
\label{m3.20p}
\FFN(\mu,h)&=\FFS(\mu,h,\Delta), \\
N n&=- \frac{\partial \FFN(\mu,h)}{\partial \mu} \label{m3.20}.
\end{align}
Together, these determine the chemical potential difference below
which the normal state is unstable to phase separation.  In our
formulation of the theory, the superfluid-state energy on the right
side of Eq.~(\ref{m3.20p}) is a minimum with respect to a variational
parameter, namely $\Delta$,
\be \frac{\partial \FFS(\mu,h,\Delta)}{\partial \Delta}=0.
\label{m3.21} \ee
Therefore, there are three parameters to solve for, namely $\Delta,
\mu$, and $h$, using these three conditions. Once these have been
determined, the upper critical polarization follows from
$P_{c2}=\frac{n_\uu-n_\dd}{n_\uu+n_\dd}=m/n$, where $m$ is the density
imbalance.
%
In the superfluid state, the single particle excitations are gapped
(for $h < \Delta$). As a consequence, at zero temperature, the
susceptibility to the polarization field $h$ vanishes, therefore, the
energy of the ground state is independent of $h$ and is given by
Eq.~(\ref{m3.10}).  In the case of the normal state, the result is
also relatively simple as the polarization field corresponds to a mere
shift of the chemical potential. The $1/\NS$ correction to the action
is
\be \STT= \sum_q
 \b^\ast(q) \Gamma^{-1}(q) \b(q)
\label{m3.23a},
\ee
where
\begin{widetext}
\begin{equation}
        \Gamma^{-1}(q) =  -\frac{1}{\lambda}+
        \frac{1}{4 V} \sum_{\bf k}  \left(
        {
         \sum_{\sigma=\pm} \tanh(\beta (\xi_{\k+\q/2}+ \sigma h)/2)+
        \tanh(\beta (\xi_{\k-\q/2}+\sigma h)/2)
        \over
        i \Omega_{\ell}-\xi_{{\bf k}+{\bf q}/2} - \xi_{{\bf k}-{\bf q}/2}
        }
        \right)
        \label{m3.23}.
\end{equation}
\end{widetext}
The free-energy density contribution due to fluctuations is given by
\be \FFNT= \frac{1}{\beta V} \sum_{q} e^{i \Omega_\ell \delta} \log
\left[ -\lambda \Gamma^{-1}(q) \right] \label{m3.24}. \ee
The zeroth order $\FFNO$ result can be obtained by taking the limit
$\Delta \to 0$ in Eq. (\ref{m2.6}) \be \FFNO= \frac{1}{V}
\sum_{\sigma=\pm} \sum_{\k}\left( \xi_\k  - \frac{2}{\beta}   \log
\left[ 2 \cosh(\beta (\xi_{\k}+\sigma h)/2) \right] \right)
\label{m3.25}. \ee

As in the preceding subsections, the parameters $\mu$ $h$, and $\Delta$
acquire corrections, due to Gaussian fluctuations, that are small in $1/\NS$,
and can therefore be written as
\begin{align}
\Delta&= \Delta\qah+ \frac{1}{\NS}\delta \Delta + \dots, \\
\mu&= \mu\qah+ \frac{1}{\NS} \delta \mu         + \dots, \\
h&= h\qah+ \frac{1}{\NS} \delta h               + \dots.
\label{m3.31}
\end{align}
Inserting this parameterization into Eqs.~(\ref{m3.20p}),
(\ref{m3.20}), and (\ref{m3.21}), we find a set of three coupled
equations giving these corrections
\begin{widetext}
\be
\begin{pmatrix}
\partial_{\mu} \FFNO-\partial_{\mu} \FFSO & \partial_{h} \FFNO - \partial_{h} \FFSO & \partial_{\Delta} \FFNO -\partial_{\Delta} \FFSO \\
\partial_{\mu \mu} \FFNO & \partial_{\mu h} \FFNO &\partial_{\mu \Delta} \FFNO \\
\partial_{\Delta \mu} \FFSO & \partial_{\Delta h } \FFSO  &\partial_{\Delta \Delta} \FFSO
\end{pmatrix}
\begin{pmatrix}
\delta \mu \\
\delta h \\
\delta \Delta
\end{pmatrix}
=-
\begin{pmatrix}
\FFNT-\FFST \\
\partial_\mu \FFNT \\
\partial_\Delta \FFST
\end{pmatrix}
\label{m3.32},
\ee
\end{widetext}
where all the derivatives are evaluated at $\Delta=\Delta\qah$,
$h=h\qah$ , and $\mu=\mu\qah$. Restricting attention to zero
temperature, many of the matrix elements vanish because : (1) at
unitarity, the superfluid state is gapped and its susceptibility to
a ``Zeeman'' field vanishes, (2) the normal state by definition does
not have any superfluid correlations, i.e., its ground-state energy
is independent of $\Delta$, (3) the ground-state energy of the
superfluid state is a minimum with respect to $\Delta$.  We then
obtain
\begin{widetext}
\be
\begin{pmatrix}
\partial_{\mu} \EENO-\partial_{\mu} \EESO & \partial_{h} \EENO  &  0  \\
\partial_{\mu \mu} \EENO & \partial_{\mu h} \EENO & 0 \\
\partial_{\Delta \mu} \EESO & 0  &\partial_{\Delta \Delta} \EESO
\end{pmatrix}
\begin{pmatrix}
\delta \mu \\
\delta h \\
\delta \Delta
\end{pmatrix}
=-
\begin{pmatrix}
\EENT-\EEST \\
\partial_\mu \EENT \\
\partial_\Delta \EEST
\end{pmatrix}
\label{m3.33}.
\ee
Solving the matrix equation (\ref{m3.33}) yields
\begin{align}
\delta \mu &= \frac{ (\EEST -\EENT)
\partial_{\mu h} \EENO +\partial_\mu \EENT
(\partial_h \EENO)}{ (\partial_\mu \EENO
-\partial_\mu \EESO )\partial_{\mu h} \EENO
-(\partial_h \EENO) \partial_{\mu \mu} \EENO } , \\
\delta h &= \frac{ -(\EEST -\EENT) \partial_{\mu \mu} \EENO
-\partial_\mu \EENT (\partial_\mu \EENO-
\partial_\mu \EESO)}{ (\partial_\mu \EENO
-\partial_\mu \EESO)
\partial_{\mu h} \EENO
-(\partial_h \EENO) \partial_{\mu \mu} \EENO } , \\
\delta \Delta &= -\frac{\partial_\Delta \EEST}{\partial_{\Delta
\Delta} \EESO} +\frac{
\partial_{\Delta \mu} \EESO}{
\partial_{\Delta \Delta} \EESO} \frac{ -(\EEST -\EENT)
\partial_{\mu h} \EENO -\partial_\mu \EENT (\partial_h \EENO)}
{ (\partial_\mu \EENO -\partial_\mu \EESO
)\partial_{\mu h} \EENO -(\partial_h \EENO) \partial_{\mu \mu} \EENO} \label{m3.37}.
\end{align}
\end{widetext}

The matrix elements, at unitarity, are
\begin{align}
\partial_{\mu} \EESO &=-1.7527 n , \\
%
%
\partial_{\mu \Delta} \EESO &=-1.5140 n/\eF, \\
\partial_{\Delta \Delta} \EESO &=1.3027 n/\eF,
\end{align}
\begin{align}
\partial_{\mu} \EENO &= -n , \\
\partial_{h} \EENO &=-0.9326 n, \\
\partial_{\mu \mu} \EENO &=- 1.2394n/\eF, \\
\partial_{\mu h} \EENO &=- 0.6290 n/\eF ,
\end{align}
\begin{align}
\partial_{h h} \EENO &=- 1.2394n/\eF , \\
\EENT &=-0.0509n\eF, \\
\EEST &=-0.477n\eF , \\
\partial_{\mu} \EENT &= -0.513n, \\
\partial_{h} \EENT &= 0.452 n, \\
\partial_{\Delta} \EEST &=-0.379n ,\\
\partial_{\mu} \EEST &=-0.948n .
\label{m3.34}
\end{align}

At this point a few remarks are in order. Some of these matrix
elements can be related to each other. Indeed, using scaling
arguments akin to those presented in Sec.~\ref{SEC:pfg}, one can show, for instance,
$\EEST=\frac{2}{5} \mu \partial_\mu \EEST +\frac{2}{5} \Delta
\partial_\Delta \EEST$, as well as $\EENT=\frac{2}{5} \mu
\partial_\mu \EENT +\frac{2}{5} h
\partial_h \EENT$.  Furthermore, we observe that the
corrections $\delta \mu$ and $\delta h$ do not involve any matrix
elements containing derivatives with respect to $\Delta$. This
merely reflects the fact that $\Delta$ is a variational parameter
and that one could have reduced the solution for $\delta \mu$ and
$\delta h$ to a set of two coupled equations using Eqs.
(\ref{m3.20}). In addition, we point out that the ratio
$\Delta/\mu$ at $h=h_{c2}$ is equal to $\Delta/\mu$ at $h=0$, a
result directly related to the vanishing susceptibility of a gapped
superfluid state.

The polarization at the upper-critical field can be written in a
$1/\NS$-expansion as
\be
P_{c2}\equiv \frac{m_{c2}}{n_{c2}}= P\qah +\frac{1}{\NS} \delta P,
\label{m3.35}
\ee
where $P\qah$ is the result of the $\NS \to \infty$
calculation and

\begin{widetext}
\be \delta P= P\qah \left[ \frac{\partial_h \EENT}{\partial_h
\EENO}
 -  \frac{\partial_\mu \EENT}{\partial_\mu \EENO}+
\left( \frac{\partial_{\mu h} \EENO}{\partial_{h} \EENO} -
\frac{\partial_{\mu\mu} \EENO}{\partial_{\mu} \EENO} \right) \delta
\mu+ \left( \frac{\partial_{hh} \EENO}{\partial_{h} \EENO} -
\frac{\partial_{\mu h} \EENO}{\partial_{\mu} \EENO} \right) \delta h
\right] \label{m3.36}. \ee

\end{widetext}
At unitarity, we find
\begin{align}
\label{m3.44a}
h_{c2} /\eF                                &=0.6930 \numbe/\NS + \curO(1/\NS^2),\\
\label{m3.44b}
P_{c2}                                &=0.9326 \numbf/\NS + \curO(1/\NS^2),\\
\label{m3.44c}
\mu_{c2}/\eF                               &=0.8586 \numbg/\NS + \curO(1/\NS^2),\\
\Delta_{c2}/\eF                            &=0.9978 \numbh/\NS +
\curO(1/\NS^2). \label{m3.44d}
\end{align}

\subsection{Away from unitarity}

Although so far we have focused on the problem at the unitarity limit,
the same large-$\NS$ expansion formalism can be used to calculate the
properties of a resonant Fermi gas away from the unitarity
point.  In this section we compute such $1/\NS$ correction at zero
temperature.

The result of Eq.~(\ref{m3.13}) holds for any value of the scattering
length $a_s$ and therefore can be used away from unitarity.  The
matrix elements of the left hand side of Eq.~(\ref{m3.13}) can be
calculated by taking appropriate derivatives of Eq.~(\ref{m2.1}).
After straightforward algebraic manipulation, one obtains
\begin{align}
\partial_{\mu \mu} \EEO &= -I_5(x_o) \frac{(2m)^{3/2}}{2 \pi^2} \Delta_o^{1/2} ,\\
\partial_{\mu \Delta} \EEO &= - I_6 (x_o)  \frac{(2m)^{3/2}}{2 \pi^2} \Delta_o^{1/2} ,\\
\partial_{\Delta \mu} \EEO &= \partial_{\mu \Delta} \EEO,  \\
\partial_{\Delta \Delta} \EEO &= I_5(x_o) \frac{(2m)^{3/2}}{2 \pi^2} \Delta_o^{1/2}.
\label{D1.1}
\end{align}
where, following Marini, \textit{ et al.} ~\cite{Marini1997},
we introduced dimensionless function $I_5(x_o)$ and $I_6(x_o)$
\begin{align}
        I_6(x_o) &= \int_0^\infty\!\!\!\! dx \frac{x^2 \xi_x }{ E_x^3}
         ,        \label{D1.2A} \\
        I_5(x_o) &= \int_0^\infty\!\!\!\! dx \frac{x^2}{E_x^3},
        \label{D1.2}
\end{align}
and dimensionless parameters
\begin{align}
x_o  &=\frac{\mu_o}{\Delta_o} \quad , &
 x^2 &=\frac{1}{\Delta_o} \frac{|\k|^2}{2m} \quad,  \notag \\
\xi_x &= \frac{\xi_\k}{\Delta_o} = x^2-x_o\quad,
& E_x &= \frac{E_\k}{\Delta_o} = \sqrt{\xi_x^2+1} .
\label{D1.3}
\end{align}

The variables $\mu_o$ and $\Delta_o$ are the solutions to the gap and particle equations in the $\NS \to \infty$ limit. For numerical purposes, it is useful to note that Eqs. (\ref{D1.2A})
and (\ref{D1.2}) can be expressed in terms of linear combinations of
complete elliptic integrals~\cite{Marini1997}.

Substituting previous equations into Eq. (\ref{m3.14}) yields,
\begin{align}
\begin{pmatrix}
\delta \mu \\
\delta \Delta
\end{pmatrix}=&
\frac{2 \eF^{3/2}}{3 n \Delta^{1/2}_o} \frac{1}{ I^2_5(x_o)+I^2_6(x_o) }
\notag \\
& \times
\begin{pmatrix}
I_5(x_o) & I_6(x_o) \\
I_6(x_o) & -I_5(x_o)
\end{pmatrix}
\begin{pmatrix}
\partial_\mu \EET \\
\partial_{\Delta} \EET
\end{pmatrix}
\label{D1.4},
\end{align}
for the leading order (in $1/N$) corrections $\delta \Delta$ and $\delta \mu$ to
the saddle-point (mean-field) result given by Eqs.~(\ref{m2.2}) and (\ref{m2.3})
at $T=0$.

Solving Eq.~(\ref{D1.4}) numerically yields results displayed
Fig.\ref{Figure1} as a function of $(k_F a_s)^{-1}$.

\section{Discussion and Comparison to other work}
\label{SECTION4}

In this work we have studied a two-component resonantly-interacting
Fermi gas in the unitary regime by generalizing the system to a
$2\NS$-component gas with $\NS$ flavors of spin-1/2 fermions,
computing a variety of thermodynamic quantities in a systematic expansion
in a small parameter $1/\NS$, with our main results presented in
Sec.~\ref{results}.  Although our results are strictly only quantitatively
valid to leading-order in $1/\NS$, to compare to experiments and other
calculations, in this section we boldly take the $\NS \to 1$ limit.

The study of the unitarity point can be seen as a benchmark for
many-body theories.  In this section, we collect the various
estimates to thermodynamic quantities at the unitarity point from
the literature and compare to our results.
  Starting with BCS expressions, a mean-field
estimate of $\xi=0.59$ can be obtained.  Since this approach is based on a
variational wavefunction, it provides a strict upper bound.  Pade
approximant techniques have also been applied to a Fermi gas
expansion in terms of $k_F \as$ to extract an estimate for
$\xi=0.326$ \cite{Heiselberg2001,Baker1999}.  Perali, \textit{ et
al.} \cite{Perali2004} have introduced a diagrammatic method based
on the t-matrix extended to the superfluid phase finding, at
unitarity, $\xi=0.455$. Most recently, Haussmann, \textit{ et
al.}~\cite{Haussmann2006} proposed a self-consistent and conserving
theory to study the crossover and found $\xi=0.36$. An epsilon
expansion of the unitarity gas, based on an expansion from the
dimension $d=4-\epsilon$ has been performed to second loop order
\cite{Arnold2006}. Depending on the Borel-Pade extrapolation schemes
used, the results range from $\xi=0.30$ to $\xi=0.37$.

At present time, the best estimates for $\xi$ coming from fixed node
Green's function Monte Carlo calculations
\cite{Carlson2003,Astrakharchik2004} yield $\xi \approx 0.44 \pm
0.01$ (by Carlson, \textit{ et al.}  \cite{Carlson2003}) and $\xi
\approx 0.42 \pm 0.01$ by Astrakharchik, \textit{ et al.}
\cite{Astrakharchik2004}. The nodal method is based on a variational
approach, thus also giving a strict upper-bound for $\xi$.

There has also been a large experimental effort aimed at extracting
the parameter $\xi$.  Experimental measurements of the expansion of
$^6$Li in the unitarity limit from a harmonic trap have determined
$\xi$ (often quoted as $\beta \equiv \xi-1$).  Experimental results
for $\xi$ were first obtained by Ghem, \textit{ et
  al.}\cite{Ghem2003}, $\xi= 0.74(7)$\ , Bartenstein, \textit{ et
  al.}\cite{Bartenstein2004} $\xi=0.32^{+10}_{-13}$ \ , and Bourdel,
\textit{ et al.}\cite{Bourdel2004}, $\xi=0.36(15)$\ . Most recently,
the Duke group \cite{Kinast2005} obtained $\xi=0.51(4)$\ , whereas the
Rice Group \cite{Partridge2006} found $\xi=0.46(5)$\ .  Experiments on
$^{40}$K by the Boulder group ~\cite{Stewart2006} yielded
$\xi=0.46^{+0.05}_{-0.12}$.
 Our prediction for $\xi = 0.28$ at $N=1$
is qualitatively consistent with these results, reflecting the smallness
of fluctuations around the saddle-point mean-field solution at $T=0$.

Another important quantity is the single particle excitation gap at unitarity.  In the
Green's function Monte Carlo calculations of
Refs.~\onlinecite{Carlson2003,Chang2004}, an estimate for the
spectroscopic energy gap $\Delta_{\rm exc}$ was determined from the odd-even
staggering of the total energy as the number of particles is increased
and found $\Delta_{\rm exc}=0.54 \eF$.  It is not a priori 
clear whether this gap is generally identical to the expectation value $\Delta\equiv |\lambda
\langle\psi_\uparrow \psi_\downarrow\rangle|$; in the
Appendix we show that they are equal to zeroth order but differ at the 
$\curO(1/N)$ computed here.  Our prediction for $\Delta_{\rm exc}/\eF$,
evaluated at $\NS=1$, is in good agreement with the Monte Carlo
calculations as seen in table \ref{Table3}.

\begin{table}[t]
\caption{Recent experimental results for $\xi$ compared with
calculated values (compilation taken from Ref.
\onlinecite{Haussmann2006} with a few additional results) }
\begin{tabular}{l | l | l}
\multicolumn{3}{l}{} \\
             &                                                     & \phantom{cent}$\xi$   \\
\hline
             & Gehm \textit{et al.}\ \cite{Ghem2003}          & $0.74(7)$ \\
             & Bartenstein \textit{et al.}\ \cite{Bartenstein2004}   & $0.32^{+0.13}_{-0.10}$ \\
Experimental & Bourdel (2004) \textit{et al.}\ \cite{Bourdel2004}    & $0.36(15)$             \\
results      & Duke (2005) \cite{Kinast2005}                         & $0.51(4)$              \\
             & Partridge \textit{et al.}\ \cite{Partridge2006}       & $0.46(5)$              \\
             & Regal \textit{et al.}\ \cite{Regal2005}            & $0.38(7)$      \\
             & Stewart \textit{et al.}\ \cite{Stewart2006}       & $0.46^{+0.05}_{-0.12}$      \\
\hline \hline
                & Mean-Field      & $\xi <0.5906$              \\
                & Astrakharchik \textit{et al.}\ \cite{Astrakharchik2004} & $0.42(1)$              \\
                & Carlson \textit{et al.}\ \cite{Carlson2003}        & $0.44(1)$              \\
Calculated      & Perali \textit{et al.}\ \cite{Perali2004}          & $0.455  $              \\
values          & Pad\'e approximation \cite{Baker1999,Heiselberg2001} & $0.33$                 \\
                & Haussmann \textit{et al.}\ \cite{Haussmann2006}      & $0.36   $              \\
                & Epsilon expansion \cite{Arnold2006}                  & $0.30 $ to $0.37$             \\
                & This work                                           & $0.28   $              \\
\end{tabular}
\label{Table1}
\end{table}

\begin{table}
\caption{Universal ratio $\Delta_{\rm exc}/\eF$ at unitarity: Comparison
between numerical and theoretical approaches.}
\begin{tabular}{l | l }
\multicolumn{2}{l}{} \\
 &  $\Delta_{\rm exc}/\eF$ \\
\hline
Mean-Field  & 0.6864\\
Carlson \textit{ et al.} (Monte Carlo) \cite{Carlson2003} & 0.54 \\
Haussmann \textit{ et al.} \cite{Haussmann2006} & 0.46 \\
Nishida \textit{ et al.} \cite{Nishida2006b}(epsilon expansion) & 0.60 \\
This work & 0.49 \\
\end{tabular}
\label{Table3}
\end{table}

The critical temperature at unitarity has also been investigated.
The $1/\NS$ calculations shows relatively large correction to the
$\NS \to \infty$ result, so that the $N\to1$ limit of Eq.~(\ref{1.3a})
is clearly not sensible {\it quantitatively\/} but can be qualitatively
interpreted as predicting a large negative correction to the mean-field
result for $T_c$.
This strong correction to the saddle-point
approximation is expected and reflects the fact that the mean-field
solution neglects the effect of bound pairs and describes the normal
state as consisting of free fermions~\cite{Nozieres1985}.
Quantum Monte calculations performed by Bulgac, \textit{ et al.}
\cite{Bulgac2006} have indeed determined a relatively low critical
temperature ($T_c=0.23(2) \eF$), and Burovski, \textit{ et al.}
\cite{Burovski2006} have arrived to $T_c=0.152(7) \eF$ using
diagrammatic determinant Monte Carlo. We note that this result is
below the BEC limit, $T_{BEC}= 0.218 \eF$, in contrast to the earlier
work of Nozieres and Schmitt-Rink \cite{Nozieres1985}. Burovski,
\textit{ et al.} also give $\mu=0.493(14)\eF$ and $\MEANE=0.31(1) \eF$
at $T_c$. Finally, we point out a recent $1/\NS$ calculation by
Nikoli\'c and Sachdev~\cite{Nikolic2006}, that gives a result for
$1/T_c$, that is consistent with our prediction for $T_c$.

Now we turn to the polarized Fermi gas. We have determined the $1/\NS$
correction to the upper critical field $P_{c2}$.  Besides the
mean-field value, we do not know of any other theoretical estimates
for $P_{c2}$. This is partly due to the difficulty of Monte Carlo
calculations to tackle the sign problem for a polarized Fermi gas.
However, a recent experiment on $^6$Li has found $P_{c2}$ to be below
its mean-field value estimate, and given by $P_{c2} =0.70$\;
\cite{Zwierlein2006}. This is consistent with the calculation of the
$1/\NS$ correction that shows that $P_{c2}$ decreases with decreasing
$\NS$ going from its mean field value $P^{\NS=\infty}_{c2}=0.933$. The
naive substitution $\NS=1$ in Eq. \ref{m3.44b} yields
$P^{\NS=1}_{c2}=0.302$, which underestimates the experimental result.

\section{Summary}
\label{SECTION5}

To summarize, we have studied a resonant Fermi gas interacting via
short-range attractive interactions by generalizing the model to
$\NS$ fermion flavors and employing a theoretical method that
is perturbative in $1/\NS$.  The $1/\NS$ expansion provides a systematic
scheme for quantitatively determining corrections to the standard BEC-BCS
mean-field theory of interacting Fermi gases.

Although our primary goal was the computation of various quantities to
leading order in $1/\NS$ in the vicinity of the unitarity point (where
universality holds), we also computed the zero-temperature gap and
chemical potential away from the unitary point.  Clearly, future work
must be done to generalize our other results for arbitrary coupling
(i.e., Feshbach resonance detuning) as well as to compute the next
$\curO (1/\NS^2)$ term.

{\em Note Added:} We thank Y. Nishida and D. T. Son for pointing out an error in an earlier version of this manuscript. 
We learned of an independent, and closely related
$1/\NS$-expansion study by P.  Nikoli\'c and S.  Sachdev
\cite{Nikolic2006}.  Where there is overlap, our results are in
agreement with theirs. 

We gratefully acknowledge useful discussions with V. Gurarie and A.
Lamacraft, as well as support from the NSF Grant DMR-0321848.

\appendix
\section{Properties of Green's functions}
\label{A}
\def\theequation{A.\arabic{equation}}%

In this Appendix we determine properties of the single-particle
excitation spectrum. In particular, we calculate the single-particle
excitation gap $\Delta_{\rm exc}$ to  subleading order in
$1/\NS$).  Although $\Delta_{\rm exc}$ (which is more commonly measured in 
experiments) is equal to the order parameter $\Delta$ within 
mean field theory (i.e., at $N\to \infty$), beyond mean-field theory they are slightly
different.  

The gap in the single-particle excitation spectrum is determined as
the lowest energy state of the single-particle Green's function,
defined as
\be
{\cal G}(x)_{\s,\si}= \langle \psipo_{1,\s}(x)
\psio_{1,\si}(0) \rangle,
 \label{A1}
\ee
where without loss of generality, we selected the flavor $i=1$.
To calculate this expectation value within our formalism, it is useful
to introduce source fields in the action
\be
S_{source} = \sum_{i=1}^{\NS}
\int_0^{\beta} d \tau \int d^3 \r \; \left[ \Etapo_{i}(x)
\Psio_{i}(x)+\Etao_{i}(x) \Psipo_{i}(x) \right], \label{A2}
\ee
where we introduced a Nambu representation for fermions, i.e.
$\Psio_{i} (x)=\left( \psio_{i\uu}(x), \psipo_{i\dd}(x) \right)$ and
where $\Etao_i(x) = \left( \etao_{i\uu}(x) ,\etapo_{i\dd}(x) \right)$ are the source fields. From the auxiliary fields, we derive
\be
{\cal G}(x)_{\s,\si} = \left.
  \frac{\partial^2  \log Z[\etao,\etapo]} {\partial \etapo_{1,\si}(0)
    \partial \etao_{1,\s} (x)} \right|_{\etao=\etapo=0}.
\label{A3}
\ee
Using the formula
\begin{align}
\int D \Psio_{i}(x) & D \Psipo_{i}(x) e^{-\sum_{ix} \left[
\Psio_i(x) M(x) \Psipo_i(x)+\Etapo_{i}(x) \Psio_{i}(x)+\Etao_{i}(x) \Psipo_{i}(x) \right]} \notag \\
&= {\rm Det}[M] e^{\sum_{ix} \Etao_{i}(x) M^{-1}(x) \Etapo_{i}(x)},
\label{A4}
\end{align}
we find the partition function is given by
\be
Z\left[ \Etao, \Etapo\right] = Z_b^{-1}\int D b(x) D b^\ast(x)
e^{-S[b]-S_{s}[\Phi]} , \label{A5}
\ee
where
\be
S_{s}[\Phi]= -\sum_{i=1}^{{\NS}} \int_0^\beta d \tau \int d^3 \r
\; \Etao_i (x) G(x) \Etapo_i (x), \label{A6}
\ee
and $S[b]$ is given by Eq. (\ref{m1.7}). Using, Eq.  (\ref{A3}), we
arrive to
%
\be {\cal G}(x)= \langle \left[ G(x) \right]_{11} \rangle_{S[b]},
\label{A7}
\ee
where the Green's function matrix is given by
\be
G^{-1}(x)=
\begin{pmatrix}
-\partial_\tau + \frac{\nabla^2}{2m} + \mu & b(x) \\
b^\ast(x)& -\partial_\tau - \frac{\nabla^2}{2m} - \mu
\end{pmatrix},
\label{A9}
\ee
Using the decomposition given in (Eq. \ref{m1.11}), we find to subleading order 
\be
\left(\langle G(k) \rangle_{S[b]}\right)^{-1} = G^{-1}_{(0)}(k)- \Sigma(k),
\ee
where 
\be
\Sigma(k)= \frac{1}{\NS} \sum_{k'} \langle G^{-1}_{(1)}(k-k') G_{(0)} (k')  G^{-1}_{(1)}(k'-k) \rangle_{S^{(1/\NS)}}.
\ee
In particular, we find for the self energy
\begin{align}
\Sigma(k)=\frac{1}{\NS} \sum_{k'} \frac{1}{|A(k-k')|^2-|B(k-k')|^2} \frac{1}{\omega'^2 + E^2_{\bf k'}} \times  \notag \\
\begin{pmatrix}
-A(k'-k) ( i \omega' - \xi_{\bf k'}) &
-B(k-k') \Delta^*  \\
-B^*(k-k') \Delta   &
-A(k-k')( i \omega' + \xi_{\bf k'})  \\
\end{pmatrix}
\end{align}

By solving the equation ${\rm det}[ \langle G  ({\bf k}, \omega)\rangle^{-1}]_{i \omega \rightarrow \omega} =0$, we obtain the dispersion relation. At leading order, i.e.
 $N \to \infty$, this is given by 
\be
\omega^{(0)}_{ \bf k}= \sqrt{\xi^2_{\bf k} +(\Delta^{(0)}_o)^2}.
\ee
Since the chemical potential is positive, it is possible to find a wavevector such that the condition $\xi_{\bf k}=0$ is fulfilled, giving, at leading order, that the lowest energy excitation is equal to the order parameter i.e. $\Delta_{\rm exc}=\Delta^{(0)}_o$. At subleading order, the additional contribution due to the self energy breaks this equality. In this case, we determine the excitation gap by expanding the dispersion relation around 
its minimum, namely $\epsilon_{\bf k}=\mu$. Straightforward manipulations give
\be
\Delta_{\rm exc}=\Delta +\frac{1}{2}  \left( \Sigma_{11}+\Sigma_{22} -2 \Sigma_{12} \right) \Big|_
{
|{\bf k}|=\sqrt{2 m \mu^{(0)}_o}, \atop i \omega=\Delta^{(0)}_o \qquad}
\ee
where $\Delta$ is the order parameter calculated at order $1/\NS$. Numerical evaluation of the self energy 
gives $\Sigma_{11}+\Sigma_{22} -2 \Sigma_{12}= -0.067 \eF$, yielding
\be
\Delta_{\rm exc}/\eF= 0.6864- 0.196/\NS + \curO(1/\NS^2),
\ee
the final result. 





\begin{thebibliography}{199}
\bibitem{narrownote}
  In the case of s-wave pairing, the typical experiments utilize
  resonances that are very broad, such that the effective range
  parameter $r_0$ satisfies $|r_0|n^{1/3} \ll 1$. It follows that the
  effective range essentially drops out of the problem, and the system
  is characterized by a single gas parameter $a_s n^{1/3}$.  As discussed
  in Refs.~\onlinecite{GRreview,ONE}, in the opposite narrow-resonance
  regime, there is a natural small parameter $\propto (k_F|r_0|)^{-1}$
  allowing a systematic perturbative study of resonantly interacting
  Fermi gases even at the unitarity point and throughout the BEC-BCS
  crossover.
\bibitem{symplectic}
To make the  $Sp(2N)$ invariance of the $N$ flavor model explicit, we write
the fermions in a column vector of length $2N$
$\Psi^T = (\psi_{1\downarrow} ... \psi_{N\downarrow}
\psi_{1\uparrow} ... \psi_{N\uparrow})$, which transforms 
Eq.~(\ref{m1.1prime}) into a form 
%
\begin{align}
H &= \int d^3 \r \,\Psi(\r)^\dagger \left( -\frac{\nabla^2_{\r}}{2 m} - \mu
\right)\Psi(\r)
\notag
\\
&+ \frac{\lambda}{4\NS} \int d^3 \r \big(
\Psi^T \Omega \Psi
\big)^\dagger \big(
\Psi^T \Omega \Psi
\big), \notag
\end{align}
%
with 
\be
\Omega =
\begin{pmatrix} 0 & \hat{I}
\\ -\hat{I} &0
\end{pmatrix},\notag
\ee
%
an antisymmetric matrix and $\hat{I}$ the $N\times N$ identity matrix.
\end{thebibliography}
\end{document}